\documentclass{sig-alternate}

\usepackage{graphicx}
\usepackage{epstopdf}
\usepackage{subfig}
\usepackage{bm}
\usepackage{enumitem}
\usepackage{url}
\usepackage[hidelinks=true]{hyperref}
\usepackage{cleveref}

\newtheorem{theorem}{Theorem}
\newtheorem{definition}{Definition}
\newtheorem{example}{Example}

\usepackage[pass]{geometry}

\def\sharedaffiliation{%
\end{tabular}
\begin{tabular}{c}}

\newfont{\eaddfntresz}{phvr8t at 11pt}

\begin{document}

\CopyrightYear{2016}
\setcopyright{rightsretained}
\conferenceinfo{Conference info n.a. (preprint)}{}
\isbn{n.a. (preprint)}
\doi{n.a. (preprint)}
\setlist[description]{font=\normalfont\itshape\space}

\title{Marciani Normal Form \\ of context-free grammars}

\numberofauthors{1}
\author{
	\alignauthor
	Giacomo Marciani\\
	\email{{\eaddfntresz gmarciani@acm.org}}
	\sharedaffiliation
	\affaddr{Department of Civil Engineering and Computer Science Engineering}  \\
	\affaddr{University of Rome Tor Vergata, Italy}
}
\maketitle

\begin{abstract}
In this paper, we prove the semidecidability of the problem of saying whether or not a
context-free grammar generates a regular language.
We introduce the notion of context-free grammar in Marciani Normal Form.
We prove that a context-free grammar in Marciani Normal Form always generates a
regular language.
\end{abstract}




\section{Introduction}
\label{sec:introduction}

In \Cref{sec:definitions}, we give some preliminary definitions about language
equations and the regularity problem.
In particular, we introduce the class of bilateral-linear language equations,
the pseudo-regular partition of productions and the looking forward property.

In \Cref{sec:marciani-rule}, we state and prove Marciani's Rule, which exposes
a method to solve the bilateral-linear language equations.

In \Cref{sec:marciani-normal-form}, we define the Marciani Normal Form, and we
prove that a context-free grammar in such a form always generates a regular
language.

In \Cref{sec:semidecidability}, we prove the semidecidability and the
undecidability of the context-free regularity problem, by the application of the 
previous results.

In \Cref{sec:examples}, we give some applicative examples of the Marciani's Rule
and the Marciani Normal Form.

\section{Definitions}
\label{sec:definitions}

We recall definitions known in the literature, for convenience of the reader.

It is well known that a language equation can be classified, at first instance,
according to the position of its unknown.
In  particular, we tipically have the following definitions.

\begin{definition}
	\label{dfn:ll-language-equation}
	A left-linear (ll) language equation in the unknown $r$ is a language
	equation of the form

	\begin{equation}
	\label{eqn:ll-language}
	r=ar+s
	\end{equation}
\end{definition}

\begin{definition}
	\label{dfn:rl-language-equation}
	A right-linear (rl) language equation in the unknown $r$ is a language
	equation of the form

	\begin{equation}
	\label{eqn:rl-language}
	r=ra+s
	\end{equation}
\end{definition}

A grammar in wich all productions for a given non-terminal identifiy ll or rl
language equations, always generates a regular language.

The problem of saying whether or not a context-free grammar $G$ generates a
regular language is very important.
In fact, a context-free grammar has greater expressiveness than a regular one.
On the other hand, a regular grammar is much lighter in terms of computational
complexity.
We call this problem \textit{Context-Free (CF) Regularity Problem}.

\begin{definition}
	\label{dfn:cf-regularity-problem}
	The \textit{Context-Free (CF) Regularity Problem} is the subset
	$\Re$ of $\{0,1\}^{*}$ defined as follows:

	\begin{equation}
	\label{eqn:cf-regularity-problem}
	\Re=\{\underline{G}|G\in[G_{CS}]\wedge L(G)\in[L_{REG}]\}
	\end{equation}

	where $\underline{G}$ is the encoding of a context-free grammar $G$ as a
	string in $\{0,1\}^{*}$, $[G_{CS}]$ is the class of context-free grammars,
	and $[L_{REG}]$ is the class of regular languages.
\end{definition}

We now introduce the new definitions, which will be essential in subsequent
sections.

\begin{definition}
	\label{dfn:bl-language-equation}
	A bilateral-linear language equation in the unknown $r$ is a language
	equation of the form

	\begin{equation}
	\label{eqn:ll-rl-language}
	r=ar+rb+s
	\end{equation}
\end{definition}

\begin{definition}
	\label{dfn:looking-forward-property}
	A grammar $G=(V_{T},V_{N},S,P)$ enjoys the looking forward property if and
	only if the digraph $D_{G}$, constructed as follows, does not have any cycle.

	Let $D_{G}$ be a digraph such that for all non-terminal symbol in $V_{T}$
	there exists a node in $D_{G}$, and for all production $A\rightarrow\alpha$
	in $P$ there exists an arc from $A$ to every non-terminal symbol in $\alpha$,
	with the exception of $A$ itself.
\end{definition}

\begin{definition}
	\label{dfn:pseudo-regular-partition}
	Given a grammar $G=(V_{T},V_{N},S,P)$, a symbol $A\in V_{N}$ and the subset
	$P_{A}\subseteq P$ of productions for the non-terminal $A$, the
	\textit{pseudo-regular partition} induced by $A$ is the partition
	$\Upsilon_{A}$ of $P_{A}$, defined as follows:

	\begin{equation}
	\label{eqn:pseudo-regular-partition}
	\Upsilon_{A}:=
	\biggl\{p\in P_{A} \;\bigg|\;
	\{A\rightarrow\alpha A\},
	\{A\rightarrow A\beta\},
	\{A\rightarrow\gamma\}
	\biggl\}
	\end{equation}

	where $\alpha,\beta,\gamma\in\left(V_{T}\cup V_{N}\setminus\{A\}\right)^{*}$.
\end{definition}

\section{Marciani's Rule}
\label{sec:marciani-rule}

It is well known that Arden's Rule permits the resolution of left-linear and
right-linear language equations \cite{Pettorossi13}.
In particular, by Arden's Rule, given the language equation $r=ar+s$ in the
unknown $r$, its least solution is $a^{*}s$. Likewise, given the language
equation $r=ra+s$ in the unknown $r$, its least solution is $sa^{*}$.

Now, we state and prove a theorem that permits the resolution of both-linear
language equations.

\begin{theorem}
	\label{thm:marciani-rule}

	Given the language equation $r=ar+rb+s$ in the unknown $r$, its least
	solution is $a^{*}sb^{*}$.

	\begin{proof}
		Let us divide the proof in the following three points.

		\begin{enumerate}[label=\alph*]

			\item We will first show that the language equation $r=ar+rb+s$ is
			equivalent to the language equation $r=a^{*}(rb+s)$.

			\item Then, we will show that $a^{*}sb^{*}$ is a solution for $r$ of
			the language equation $r=a^{*}(rb+s)$.

			\item Finally, we will show that $a^{*}sb^{*}$ is the minimal
			solution for $r$ of the language equation $r=a^{*}(rb+s)$; that is,
			for	any other solution $z$ we have that
			$L(a^{*}sb^{*})\subseteq L(z)$.

		\end{enumerate}

		\begin{description}

			\item [Proof (a)] By the application of Arden's Rule on $r=ar+rb+s$, we obtain $r=a^{*}(rb+s)$.
			So $ar+rb+s=a^{*}(rb+s)$\cite{Pettorossi13}.

			\item [Proof (b)] Notice that $a^{*}((a^{*}sb^{*})b+s)=a^{*}a^{*}sb^{*}b+a^{*}s=a^{*}sb^{*}b+a^{*}s$,
			so we have to show that
			(b.1) $a^{*}sb^{*}\subseteq a^{*}sb^{*}b+a^{*}s$ and
			(b.2) $\beta.2$ $a^{*}sb^{*}b+a^{*}s\subseteq a^{*}sb^{*}$.

			\begin{description}

				\item [Proof (b.1)] The following inclusion holds
				$a^{*}sb^{*}=a^{*}s(b^{+}+\varepsilon)\subseteq  a^{*}s(b^{*}b+\varepsilon)=a^{*}sb^{*}b+a^{*}s$.

				\item [Proof (b.2)] The following inclusions hold
				$a^{*}sb^{*}b\subseteq a^{*}sb^{*}$ and $a^{*}s\subseteq a^{*}sb^{*}$.

			\end{description}

			\item [Proof (c)] We assume that $z$ is a solution of $r=a^{*}(rb+s)$, that is $z=a^{*}(zb+s)$, and we show that $a^{*}sb^{*}\subseteq z$,
			that is $\bigcup_{i,j\geq0}a^{i}sb^{j}\subseteq z$. The proof can be done by induction on $i,j\geq0$.
			
			\begin{description}
				
				\item [(Basis:$i,j=0$)] $s\subseteq z$ holds because
				$z=a^{*}(zb+s)$.
				
				\item [(Step:$i\geq0,j=0$)] We have to show the following implication
				$\bigcup_{i\geq0}a^{i}s\subseteq z\rightarrow\bigcup_{i\geq0}a^{i+1}s\subseteq z$.
				This holds because $\bigcup_{i\geq0}a^{i+1}s\subseteq a\bigcup_{i\geq0}a^{i}s\subseteq az\subseteq z$.
				
				\item [(Step:$i,j\geq0$)] We have to show the following implication
				$\bigcup_{i,j\geq0}a^{i}sb^{j}\subseteq z\rightarrow\bigcup_{i,j\geq0}a^{i}sb^{j+1}\subseteq z$.
				This holds because $\bigcup_{i,j\geq0}a^{i}sb^{j+1}\subseteq\bigcup_{i,j\geq0}a^{i}sb^{j}b\subseteq zb\subseteq z$.
				
			\end{description}

		\end{description}		
	\end{proof}
\end{theorem}

\section{Marciani Normal Form}
\label{sec:marciani-normal-form}

We introduce the notion of context-free grammars in Marciani Normal Form (MNF).
We prove that every MNF grammar generates a regular language.

\begin{definition}
	\label{dfn:mnf}
	A context-free grammar $G=(V,V_{N},S,P)$ is said to be in Marciani Normal
	Form (MNF) iff $G$ enjoys the looking forward property and there exists a
	pseudo-regular partition $\Upsilon_{A}$ for all symbol $A\in V_{T}$.
\end{definition}

\begin{theorem}
	\label{thm:mnf}
	A context-free grammar $G=(V_{T},V_{N},S,P)$ in Marciani Normal Form always
	generates a regular language.

	\begin{proof}
		If $G$ is in Marciani Normal Form, then for each non-terminal $A$ there
		exists a pseudo-regular partition $\Upsilon_{A}$.
		Notice that to every pseudo-regular partition
		$\Upsilon_{A}=\{\{p|p\in P_{A},A\rightarrow\alpha A\},
		\{p|p\in P_{A},A\rightarrow A\beta\},\{p|p\in P_{A},A\rightarrow\gamma\}\}$
		can be associated a both-linear language equation $A=\alpha A+A\beta+\gamma$.
		By the application of the Marciani's Rule, we know that the least
		solution of the previous equation is $\alpha^{*}\gamma\beta^{*}$,
		so $L(A)=L(\alpha^{*}\gamma\beta^{*})$.

		As a consequence of the previous results and the looking forward property,
		there exists a regular expression $e$ such that $L(S)=L(e)$, then $L(S)$ 
		is a regular language, that is $G$ is regular.
	\end{proof}
\end{theorem}

\section{Semidecidability of the CF Regularity Problem}
\label{sec:semidecidability}

We know that the context-free regularity problem is undecidable
\cite{Pettorossi13}, due to its reduction to the undecidable Post Correspondence
Problem \cite{Hopcroft06}.

We prove the semidecidability and the undecidability of the context-free
regularity problem.

\begin{theorem}
	\label{thm:semidecidability-mnf}
	Given a context-free grammar, the problem of saying whether or not there
	exists an equivalent grammar in Marciani Normal Form is semidecidable and
	undecidable.

	\begin{proof}
		Let us consider a context-free grammar in Chomsky Normal Form.
		Let us derive an equivalent grammar by unfolding every production,
		until getting the productions for the axiom only.
		Now, it' easy to check if the derived grammar is in Marciani Normal Form.
		So, given any context-free grammar in Chomsky Normal Form, it is always
		possible to check if there exists an equivalent context-free grammar in
		Marciani Normal Form.

		As this possibility holds for grammars in Chomsky Normal Form, then it
		holds for every context-free grammar \cite{Pettorossi13}.
	\end{proof}
\end{theorem}

\begin{theorem}
	\label{thm:semidecidability}
	The context-free regularity problem is semidecidable and undecidable.

	\begin{proof}
		Follows from the semidecidability and undecidability of the problem of
		saying whether or not, given a context-free grammar, there exists an
		equivalent grammar in Marciani Normal Form, and from the regularity of
		the language generated by a context-free grammar in Marciani Normal Form.
	\end{proof}
\end{theorem}

\section{Examples}
\label{sec:examples}

We now see three applicative examples of the Marciani Normal Form and Marciani's
Rule.
Let us first consider a very simple application.

\begin{example}
	Let us consider the context-free grammar $G$ with axiom $S$ and
	the following productions

	\begin{flalign*}
		S&\rightarrow abcS|Sdef|ghi|\varepsilon
	\end{flalign*}

	The grammar is in MNF, thus we know it generates a context-free language.
	The language generated by $G$ is denoted by the regular expression

	\begin{equation*}
		(abc)^{*}(ghi+\varepsilon)(def)^{*}
	\end{equation*}
\end{example}

Let us now consider a slightly more complex application.

\begin{example}
	Let us consider the context-free grammar $G$ with axiom $S$ and
	the following productions

	\begin{flalign*}
		S&\rightarrow aAS|SBdef|CD|\varepsilon \\
		A&\rightarrow uA|Av|m \\
		B&\rightarrow xB|By|n \\
		C&\rightarrow gC|Ch|i \\
		D&\rightarrow pD|Dq|r
	\end{flalign*}

	The grammar is in MNF, thus we know it generates a context-free language.
	The language generated by $G$ is denoted by the regular expression

	\begin{equation*}
		(au^{*}mv^{*})^{*}(g^{*}ih^{*}p^{*}rq^{*}+\varepsilon)(x^{*}ny^{*}def)^{*}
	\end{equation*}
\end{example}

We now consider an application on a grammar in the notable Chomsky Normal Form
(CNF) \cite{chomsky1959certain}.

\begin{example}
	Let us consider the context-free grammar $G$ with axiom $S$ and the following
	productions

	\begin{flalign*}
		S&\rightarrow AS|SB|CD|z|\varepsilon \\
		A&\rightarrow UA|AU|m \\
		B&\rightarrow XB|BX|n \\
		C&\rightarrow UC|CU|i \\
		D&\rightarrow XD|DX|r \\
		U&\rightarrow u \\
		X&\rightarrow x
	\end{flalign*}

	The grammar is in MNF, thus we know it will generate a context-free language.
	The language generated by $G$ is denoted by the regular expression

	\begin{equation*}
		(u^{*}mu^{*})^{*}(u^{*}iu^{*}x^{*}rx^{*}+z+\varepsilon)(x^{*}nx^{*})^{*}
	\end{equation*}
\end{example}

\section{Conclusions}
\label{sec:conclusions}

We introduced the notion of context-free grammar in Marciani Normal Form (MNF).
We proved that a context-free grammar in such a form always generates a regular
language.
Such a demonstration states the semidecidability of the CF Regularity Problem,
which is the problem of saying whether or not a context-free grammar generates a
regular language.
We gave representative examples of applying Marciani's Rule, showing the
simplicity of determining the solution for the CF Regularity Problem.

\bibliographystyle{abbrv}
\bibliography{./ref/biblio}

\end{document}